\documentclass[aps,twocolumn,english,10pt,nofootinbib,superscriptaddress,showkeys,showpacs,pra]{revtex4-1}
\usepackage[LGR,T1]{fontenc}
\usepackage[latin9]{inputenc}
\usepackage{color,times}
\usepackage{amsmath}
\usepackage{amssymb}
\usepackage{graphicx}
\usepackage{ulem}

\makeatletter

\DeclareRobustCommand{\greektext}{%
  \fontencoding{LGR}\selectfont\def\encodingdefault{LGR}}
\DeclareRobustCommand{\textgreek}[1]{\leavevmode{\greektext #1}}
\ProvideTextCommand{\~}{LGR}[1]{\char126#1}

\usepackage[caption=false]{subfig}

\usepackage{babel}
\begin{document}
\title{Interplay between quantum Zeno and anti-Zeno effects in a non-degenerate hyper-Raman nonlinear optical coupler}
	
	\author{Moumita Das}
	\affiliation{Department of Physics, Siliguri College, Siliguri - 734 001, India}	
		
	\author{Kishore Thapliyal}
	\email{kishore.thapliyal@upol.cz}
	\affiliation{RCPTM, Joint Laboratory of Optics of Palacky University
and Institute of Physics of Academy of Science of the Czech Republic,
Faculty of Science, Palacky University, 17. listopadu 12, 771 46 Olomouc,
Czech Republic}
	
	\author{Biswajit Sen}
	\email{bsen75@yahoo.co.in}
	\affiliation{Department of Physics, Vidyasagar Teachers' Training College, Midnapore - 721 101, India }
	
	\author{Jan Pe{\v r}ina}
	\affiliation{Joint Laboratory of Optics of Palacky University
and Institute of Physics of Academy of Science of the Czech Republic,
Faculty of Science, Palacky University, 17. listopadu 12, 771 46 Olomouc,
Czech Republic}
	
	\author{Anirban Pathak}
	\email{anirban.pathak@jiit.ac.in}
	\affiliation{Jaypee Institute of Information Technology, A-10, Sector-62, Noida UP-201309, India}

\begin{abstract}
Quantum Zeno and anti-Zeno effects are studied in an asymmetric nonlinear
optical coupler composed of a probe waveguide and a system waveguide. The system
is a nonlinear waveguide operating under non-degenerate hyper-Raman
process, while both the pump modes in the system are constantly interacting
with the probe waveguide. The effect of the presence of probe on the
temporal evolution of the system in terms of the number of photons
in Stokes and anti-Stokes modes as well as phonon number is quantified
as Zeno parameter. The negative (positive) values of the Zeno parameter
in the specific mode are considered as the signatures of the quantum Zeno (anti-Zeno)
effect in that mode of the system. It is observed that the phase mismatch in Stokes
and anti-Stokes generation processes can be controlled to induce a
transition between quantum Zeno and anti-Zeno effects for both off-resonant
and resonant hyper-Raman process. However, in case of off-resonant
hyper-Raman process in the system waveguide, the frequency detuning
parameters can also be used analogously to cause the desired crossover. Further, the general nature of the physical system and the perturbative technique used here allowed us to analytically study the possibilities of observing quantum Zeno and anti-Zeno effects in a large number of special cases, including situations where the process is spontaneous, partially spontaneous and/or the system
is operated under degenerate hyper-Raman process, or a simple Raman process.
\end{abstract}
\maketitle

\section{Introduction}

The response of a quantum system to a measuring device, cannot only
distinctively distinguish a quantum system from a classical system,
it also plays an extremely important role in the subsequent evolution
of the system. Interestingly, sufficiently frequent interactions can
even suppress the time evolution. This phenomenon of suppressing (inhibition
of) the time evolution of a quantum system by frequent interaction
is known as the quantum Zeno effect (QZE) \cite{misra1977zeno}.
It was introduced by Mishra and Sudarshan in 1977\textcolor{black}{
\cite{misra1977zeno}, in an interesting work, where they showed
that if an unstable particle is continuously measured, it will never
decay. They realized that this situation is analogous to one of the
}\textcolor{black}{well-known
Zeno's paradoxes which were introduced by the Greek philosopher Zeno
of Elea in the 5th century BC and which have persistently
fascinated scientists and philosophers since then. Considering the
analogy, Mishra and Sudarshan referred to the quantum phenomenon analogous
to classical Zeno's paradox as } \textit{Zeno's paradox in quantum theory}.
{This led to the formal origin of QZE, but the quantum analogue of Zeno's paradox
was also studied before the work of Mishra and Sudarshan \cite{misra1977zeno}.
Specifically, Khalfin had studied nonexponential decay of unstable
atoms \cite{Khalfin} in the late fifties and early sixties. Interestingly,
it was soon recognized that measurement can also lead to a phenomenon
which can be viewed as the inverse of QZE. In such a phenomenon, continuous
measurement (or interaction) leads to the enhancement of the evolution instead of the inhibition
(see Refs. \cite{venugopalan2007zeno_review,facchi2001zeno-review,saverio-zeno-in-70-minutes,chi2-chi1-spie,NZeno,PT})
and is referred to as quantum anti-Zeno effect (QAZE) or inverse Zeno
effect. Interestingly, QZE and QAZE are known to be evinced through
various equivalent ways \cite{saverio-3-manifestation}. For the
present work, a particularly relevant manifestation of QZE would be
one in which QZE or QAZE is viewed as a process led by continuous
interaction between a system and a probe. Specifically, in what follows,
we aim to study the continuous interaction-type manifestation of QZE
in hyper-Raman processes, where it will be considered that a nonlinear
waveguide is operating under non-degenerate hyper-Raman process and
is continuously interacting with a probe waveguide. A change in the
dynamics of the nonlinear system waveguide due to the presence of
the probe waveguide is quantified as increase/decrease in the photon
numbers of Stokes and anti-Stokes modes as well as phonon number.
Earlier, QZE is reported by some of the present authors in Raman process
\cite{thun2002zeno-raman}, an asymmetric and a symmetric nonlinear
optical couplers \cite{chi2-chi1-spie,NZeno}, and parity-time symmetric
linear optical coupler \cite{PT}. However, it }{was
never been } {studied in the systems involving hyper-Raman
process due to its intrinsic mathematical difficulty. }

{Apart from the Raman process, QZE and QAZE has already
been studied in various optical systems, like various types of optical
couplers \cite{Rechacek-2001-zrno-coupler,thun2002zeno-raman,chi2-chi1-spie,chi2-chi2,rehacek2000zeno-coupler},
parametric down-conversion \cite{down-conversion-perina2,parametric-down-conversion-anti-zeno,down-conversion-perina},
parametric down conversion with losses \cite{perina-zeno-parametric-dc},
an arrangement of beam splitters \cite{All-optical-zeno-agarwal}.
In these studies on QZE in optical systems, often the pump mode has
been considered strong, and thus the complexity of a completely quantum
mechanical treatment has been circumvented. Keeping this in mind,
here we plan to use a completely quantum mechanical description of
the} nonlinear optical coupler composed of a probe and a nonlinear
waveguide operating under  hyper-Raman process. We have considered
a general Raman process here, namely non-degenerate hyper-Raman process,
which allows us to reduce the corresponding results for Raman and
degenerate hyper-Raman processes.

\begin{figure*}
\begin{centering}
\includegraphics[scale=0.7]{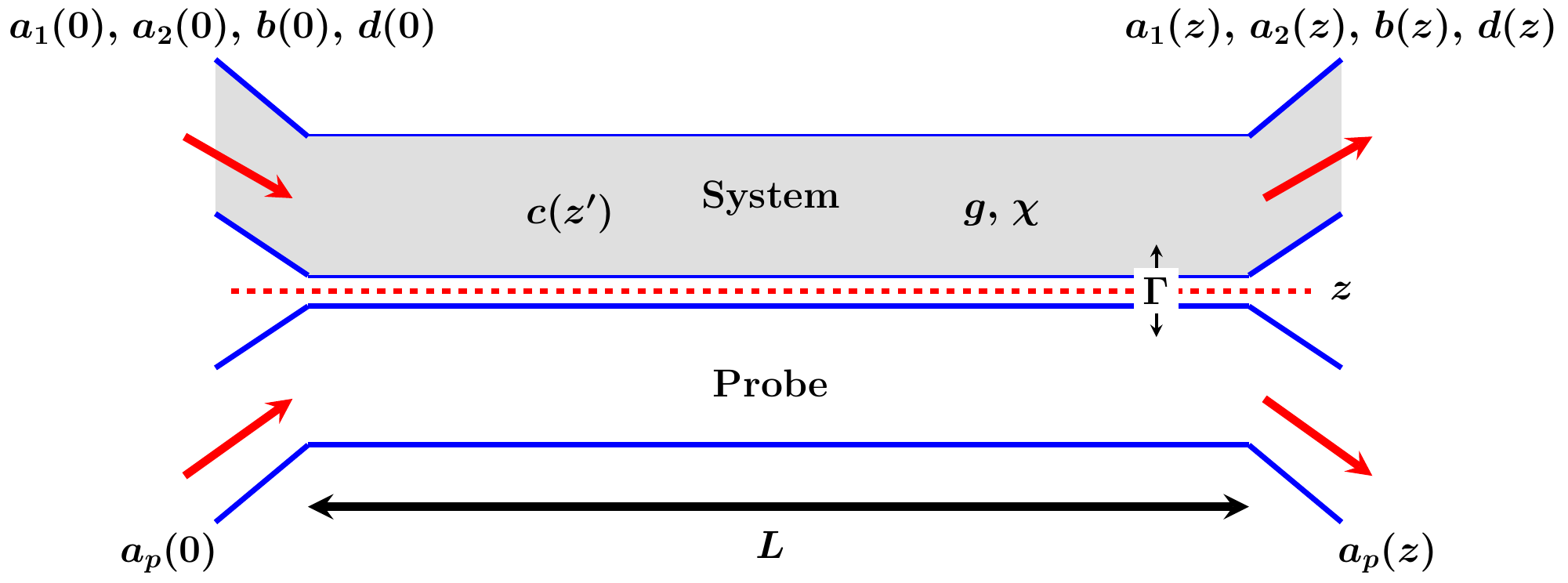}
\par\end{centering}
\caption{\label{fig:physical-system}(Color online) Schematic diagram of an
optical coupler of length $L$ composed a system waveguide, operating
under non-degenerate hyper-Raman process, interacting with a probe.
The coupling coefficients as well as optical and phonon modes are
indicated in the diagram.}
\end{figure*}

For about two decades, since the pioneering work of Mishra and Sudarashan,
QZE remained as a problem of theoretical interest without any practical
applications. The scenario started changing from the beginning of
nineties as it became possible to experimentally realize QZE using
different routes {{} \cite{kwait-int.-free.measurement-expt.,experimental-zeno-1,experimental-zeno-2}.
}\textcolor{black}{This enhanced the interest on QZE and that in turn
led to many new proposals for applications of QZE \cite{kwait-int.-free.measurement-expt.,counterfactual-quantum-computation,Zubairy1,zeno-tomography-hradil,zeno-tomography2}.
For example, applications of QZE were proposed for the enhancement
of the resolution of absorption tomography \cite{zeno-tomography-hradil,zeno-tomography2},
reduction of communication complexity \cite{Q-com-comp}, in combating
decoherence by confining dynamics in decoherence-free subspace \cite{Decoherence-free}.
Further, proposals for quantum interrogation measurement \cite{kwait-int.-free.measurement-expt.}
and counterfactual direct quantum communication protocol \cite{Zubairy1,coun-comEx}
using QZE have garnered much attention. Efforts have also been made to
investigate QZE in the macroscopic world for large black holes \cite{Zeno-blackhole}
and in nonlinear waveguides in the context of stationary flows with
localized dissipation \cite{macroscopic-zenoPRL}. Inspired by these
applications of QZE, a general nature of the nonlinear (non-degenerate
hyper-Raman) process under consideration, and availability of nonlinear
optical couplers in integrated optics and optical fiber \cite{obrien1,obrien2},
 here we study the possibility of QZE and QAZE using a completely quantum
treatment in an asymmetric nonlinear optical coupler.}

\textcolor{black}{Using a perturbative technique (known as Sen-Mandal
technique \cite{kishore2014co-coupler,kishore2014contra,mandal2004co-coupler,bsen1})
for obtaining an operator solution of Heisenberg's equations of motion,
we have obtained closed form analytic expressions for the spatial
evolution of the relevant field operators present in the system momentum
operator that provides a completely quantum mechanical description of}
nonlinear optical coupler composed of the probe and a nonlinear waveguide
operating under hyper-Raman process. {It is well
established that this method produces better results compared to the
conventional short-length approach \cite{kishore2014co-coupler,kishore2014contra,mandal2004co-coupler,bsen1}
as the present solution is not restricted by length/time.} In what follows, we will see that the use of this perturbative technique has revealed compact analytic expressions for Zeno parameter which clearly illustrate that in the system under consideration (and in many special cases of it), it is possible to observe QZE and QAZE in Stokes, anti-Stokes, and phonon modes for the suitable choices of system parameters. Further, it will be shown that phase mismatch and frequency detuning parameters can be controlled to execute a crossover between QZE and QAZE.

{The rest of the paper is organized as follows. In
Section \ref{sec:physical-system}, we briefly describe the hyper-Raman
process based asymmetric nonlinear optical coupler system of our interest.
Subsequently, we report the expressions of Zeno parameter for both
photon and phonon modes in system in Section \ref{sec:Quantum-zeno-and}.
A detailed discussion of the obtained results is summarized in Section
\ref{sec:Discussion}. Finally, the paper is concluded in Section
\ref{sec:Conclusion}.}

\section{physical system \label{sec:physical-system}}

The physical system of our interest is actually a codirectional asymmetric
nonlinear optical coupler composed of a probe and a system, which
is a nonlinear waveguide operating under hyper-Raman process (see
Fig. \ref{fig:physical-system} for a schematic diagram). The momentum
operator of the complete (probe+system) physical system is given by
\begin{equation}
\begin{array}{lcl}
G & = & \omega_{p}a_{p}^{\dagger}a_{p}+\omega_{a_{1}}a_{1}^{\dagger}a_{1}+\omega_{a_{2}}a_{2}^{\dagger}a_{2}+\omega_{b}b^{\dagger}b+\omega_{c}c^{\dagger}c+\omega_{d}\\
&\times&d^{\dagger}d
+\left(ga_{1}a_{2}b^{\dagger}c^{\dagger}+\chi a_{1}a_{2}cd^{\dagger}+\Gamma a_{p}a_{1}^{\dagger}a_{2}^{\dagger}+{\rm H.c.}\right),\end{array}\label{eq:ham}
\end{equation}
\textcolor{black}{where $\hbar=1$ (the same convention is used in
the rest of the paper), and H.c. stands for the Hermitian conjugate. }{The
annihilation (creation) operators $a_{p}(a_{p}^{\dagger}),\,a_{i}(a_{i}^{\dagger}),\,b(b^{\dagger}),\,c\left(c^{\dagger}\right)$, and $d(d^{\dagger})$
correspond to the probe pump (indexed by subscript $p)$, non-degenerate
hyper-Raman laser (this is also a pump, but it is indexed by subscript
$i=1,2$ to distinguish from the probe pump), Stokes, vibration (phonon),
and anti-Stokes modes, respectively. All field operators introduced
here, obey the usual bosonic commutation relation. Frequencies corresponding
to the probe, Pump 1 and 2, Stokes, phonon and anti-Stokes modes 
are denoted by $\omega_{p},$ $\omega_{a_{1}}$, $\omega_{a_{2}},$ $\omega_{b}$,
$\omega_{c}$, and $\omega_{d}$, respectively.} The Stokes and anti-Stokes
coupling constants are described by parameters $g$ and $\chi$, respectively.\textcolor{red}{{}
}Further, $\Gamma$ denotes the interaction constant between the probe
and the two pump modes present in the hyper-Raman process in the system.
We obtain the Heisenberg's equations of motion for the momentum operator
(\ref{eq:ham}) of the system of interest, which gives us six coupled
differential equations for the six bosonic operators present in the expression of the momentum operator. To obtain the spatial
evolution of all the operators we use Sen-Mandal perturbative technique
as dimensionless quantities $gz,$ $\chi z$, and $\Gamma z$ are
small compared to unity. In Appendix
\ref{sec:app-A}, we have reported the analytic operator  solution of the Heisenberg's equations
of motion using the Sen-Mandal technique and spatial evolution of the relevant field operators including number operators for Stokes, anti-Stokes, and phonon modes. In the next section, we will show that the obtained number operators would provide us analytic expressions of Zeno parameter for the respective modes.  The general nature of the solution and the process
under consideration allow us to deduce the results for Raman and degenerate
hyper-Raman processes as well as corresponding short-length solution in the limiting cases.
Further, the present solution inherently introduces frequency detuning
parameters in the Stokes and anti-Stokes generation, and thus provides
solutions for both resonant and off-resonant set of Raman processes.

\section{Quantum zeno and anti-zeno effect \label{sec:Quantum-zeno-and}}

In order to investigate the QZE and QAZE in hyper-Raman active medium
we consider the initial composite coherent state $\left|\psi(0)\right\rangle $
as the product of the initial coherent states of probe, pump, Stokes,
phonon and anti-Stokes modes as $\left|\alpha\right\rangle $, $\left|\alpha_{j}\right\rangle $,
$\left|\beta\right\rangle $, $\left|\gamma\right\rangle $, and $\left|\delta\right\rangle $,
respectively. Hence the initial state is considered to be
\begin{equation}
\begin{array}{lcl}
\left|\psi(0)\right\rangle  & = & \left|\alpha\right\rangle \otimes\left|\alpha_{1}\right\rangle \otimes\left|\alpha_{2}\right\rangle \otimes\left|\beta\right\rangle \otimes\left|\gamma\right\rangle \otimes\left|\delta\right\rangle ,\end{array}\label{eq:inSt}
\end{equation}
and the field operator $a_{p}$ operating on the initial state gives
\begin{equation}
\begin{array}{lcl}
a_{p}(0)\left|\psi(0)\right\rangle  & =\alpha & \left|\alpha\right\rangle \otimes\left|\alpha_{1}\right\rangle \otimes\left|\alpha_{2}\right\rangle \otimes\left|\beta\right\rangle \otimes\left|\gamma\right\rangle \otimes\left|\delta\right\rangle ,\end{array}\label{eq:ann}
\end{equation}
where $\begin{array}{ccc}
\alpha & = & \left|\alpha\right|e^{i\varphi_{p}}\end{array}$ is the complex eigenvalue with $\left|\alpha\right|^{2}$ mean number
of photons and $\varphi_{p}$ phase angle in the probe mode $a.$
In the similar manner, coherent state parameters
for all the optical and phonon modes involved in the hyper-Raman process
can be defined as $\Lambda_{j}=\left|\Lambda_{j}\right|e^{i\varphi_{j}}$
for complex amplitudes $\Lambda_{j}:\Lambda\in\left\{ \alpha,\beta,\gamma,\delta\right\} $
and corresponding phase angle $\varphi_{j}$ with $j\in\left\{ 1,2,b,c,d\right\} $
for the non-degenerate Pump-1 and Pump-2, Stokes, phonon, and anti-Stokes
modes, respectively.

We define the Zeno parameter as \cite{chi2-chi1-spie,NZeno,PT}\textcolor{blue}{{}
}
\begin{equation}
\begin{array}{lcl}
Z_{i} & = & \left\langle N_{i}\right\rangle -\left\langle N_{i}\right\rangle _{\Gamma=0},\end{array}\label{eq:Z}
\end{equation}
where $i\in\left\{ b,c,d\right\} $ and $N$ is the number operator.
The conditions $Z_{i}<0$ and $Z_{i}>0$ correspond to the occurrence
of QZE and QAZE in $i$th mode. Clearly, the negative (positive) values
of the Zeno parameter represent that the number of photon/phonons
 in the system waveguide (i.e., $\left\langle N_{i}\right\rangle $) decreases
(increases) from that in the
absence of the probe waveguide  (i.e., $\left\langle N_{i}\right\rangle _{\Gamma=0}$). We have reported the dynamics of number
operators for Stokes, anti-Stokes, and phonon modes in Appendix \ref{sec:app-A}.
In what follows, we report the Zeno parameters for Stokes, anti-Stokes,
and phonon modes obtained using these expressions.

To begin with, using Eq. (\ref{eq:nb}) in Eq. (\ref{eq:Z}), the
Zeno parameter for the Stokes mode is computed as
\begin{equation}
\begin{array}{lcl}
Z_{b} & = & \mathcal{C}_{b}\left\{ \frac{\cos\theta_{2}}{\text{\textgreek{D}}\omega_{S}\left(\text{\textgreek{D}}\omega_{D}+\text{\textgreek{D}}\omega_{S}\right)}+\frac{\cos\left(\text{\textgreek{D}}\omega_{S}z+\text{\textgreek{D}}\omega_{D}z-\theta_{2}\right)}{\text{\textgreek{D}}\omega_{D}\left(\text{\textgreek{D}}\omega_{D}+\text{\textgreek{D}}\omega_{S}\right)}\right.\\
&-&\left.\frac{\cos\left(\text{\textgreek{D}}\omega_{S}z-\theta_{2}\right)}{\text{\textgreek{D}}\omega_{D}\text{\textgreek{D}}\omega_{S}}\right\} ,\end{array}\label{eq:Zb}
\end{equation}
where we have used phase mismatches with probe--Stokes and probe--anti-Stokes
process as $\theta_{1}=\left(\varphi_{d}-\varphi_{p}-\varphi_{c}\right)$
and $\theta_{2}=\left(\varphi_{p}-\varphi_{b}-\varphi_{c}\right)$, respectively.
Also, we have $\mathcal{C}_{b}=2\Gamma g\left(\left|\alpha_{1}\right|^{2}+\left|\alpha_{2}\right|^{2}+1\right)\left|\alpha\right|\left|\beta\right|\left|\gamma\right|$.
Similarly, we have introduced frequency detuning parameters in Stokes,
anti-Stokes, and second-order coupling between the probe and system waveguides
as $\Delta\omega_{S}=-\omega_{a_{1}}-\omega_{a_{2}}+\omega_{b}+\omega_{c}$,
$\Delta\omega_{A}=\omega_{a_{1}}+\omega_{a_{2}}+\omega_{c}-\omega_{d}$,
and $\Delta\omega_{D}=\omega_{a_{1}}+\omega_{a_{2}}-\omega_{p}$. 

In the similar manner, the Zeno parameter for the anti-Stokes mode
is computed using Eqs. (\ref{eq:Z}) and (\ref{eq:nd}) and the following
analytic expression is obtained
\begin{equation}
\begin{array}{lcl}
Z_{d} & = & \mathcal{C}_{d}\left\{ \frac{\cos\theta_{1}}{\text{\textgreek{D}}\omega_{A}\left(\text{\textgreek{D}}\omega_{A}-\text{\textgreek{D}}\omega_{D}\right)}-\frac{\cos\left(\theta_{1}+\text{\textgreek{D}}\omega_{D}z-\text{\textgreek{D}}\omega_{A}z\right)}{\text{\textgreek{D}}\omega_{D}\left(\text{\textgreek{D}}\omega_{A}-\text{\textgreek{D}}\omega_{D}\right)}\right.\\
&+&\left.\frac{\cos\left(\theta_{1}-\text{\textgreek{D}}\omega_{A}z\right)}{\text{\textgreek{D}}\omega_{D}\text{\textgreek{D}}\omega_{A}}\right\} ,\end{array}\label{eq:Zd}
\end{equation}
where $\mathcal{C}_{d}=2\Gamma\chi\left(\left|\alpha_{1}\right|^{2}+\left|\alpha_{2}\right|^{2}+1\right)\left|\alpha\right|\left|\gamma\right|\left|\delta\right|$.
Zeno parameter for the phonon mode was also computed analytically
using Eqs. (\ref{eq:Z})-(\ref{eq:nc}), but it was found that the
same can be expressed as the difference of the other two Zeno parameters,
i.e.,
\begin{equation}
\begin{array}{lcl}
Z_{c} & = & Z_{b}-Z_{d}.\end{array}\label{eq:Zc}
\end{equation}
This is a direct consequence of the conservation law: $\left[G,N_{c}+N_{d}-N_{b}\right]=0$,
and thus using this constant of motion \cite{perina-book} Eq. (\ref{eq:Zc})
follows.  Apparently, this should be the case with
degenerate hyper-Raman and Raman processes in the system, too, provided
the probe-system interaction is with the pump mode, however, the probe-system
interaction may be with the Stokes and anti-Stokes modes as in the
case of earlier works \cite{thun2002zeno-raman}.

\begin{figure*}
\begin{centering}
\subfloat[]{\begin{centering}
\includegraphics[scale=0.6]{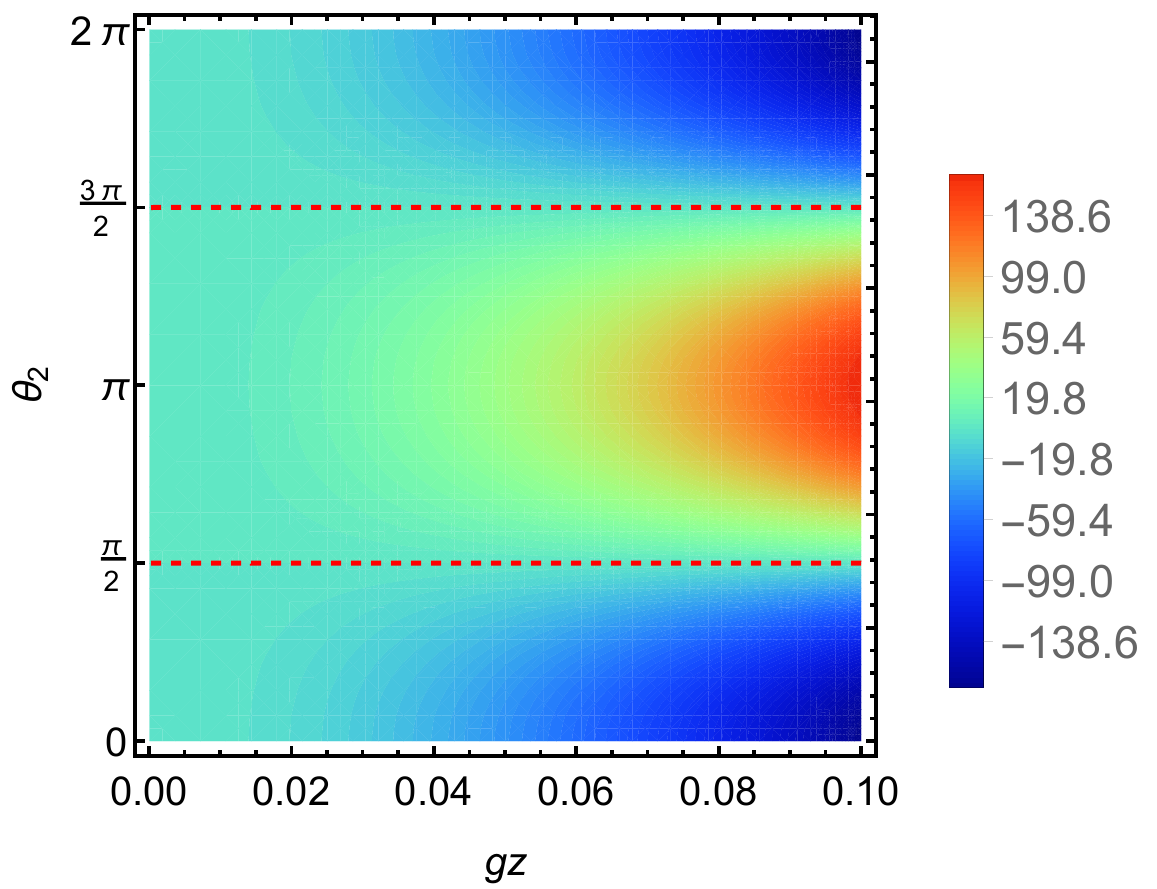} 
\par\end{centering}
} \subfloat[]{\begin{centering}
\includegraphics[scale=0.6]{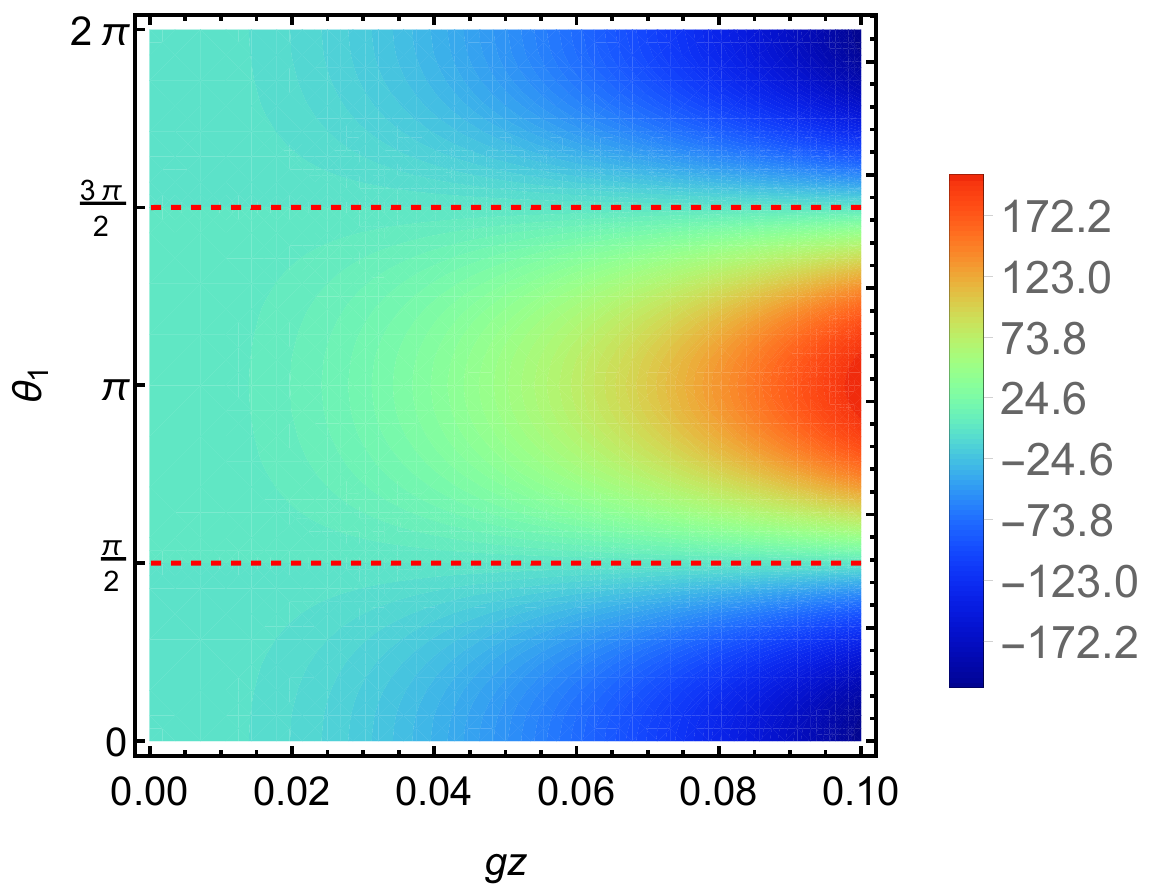}
\par\end{centering}
}
\par\end{centering}
\begin{centering}
\subfloat[]{\begin{centering}
\includegraphics[scale=0.6]{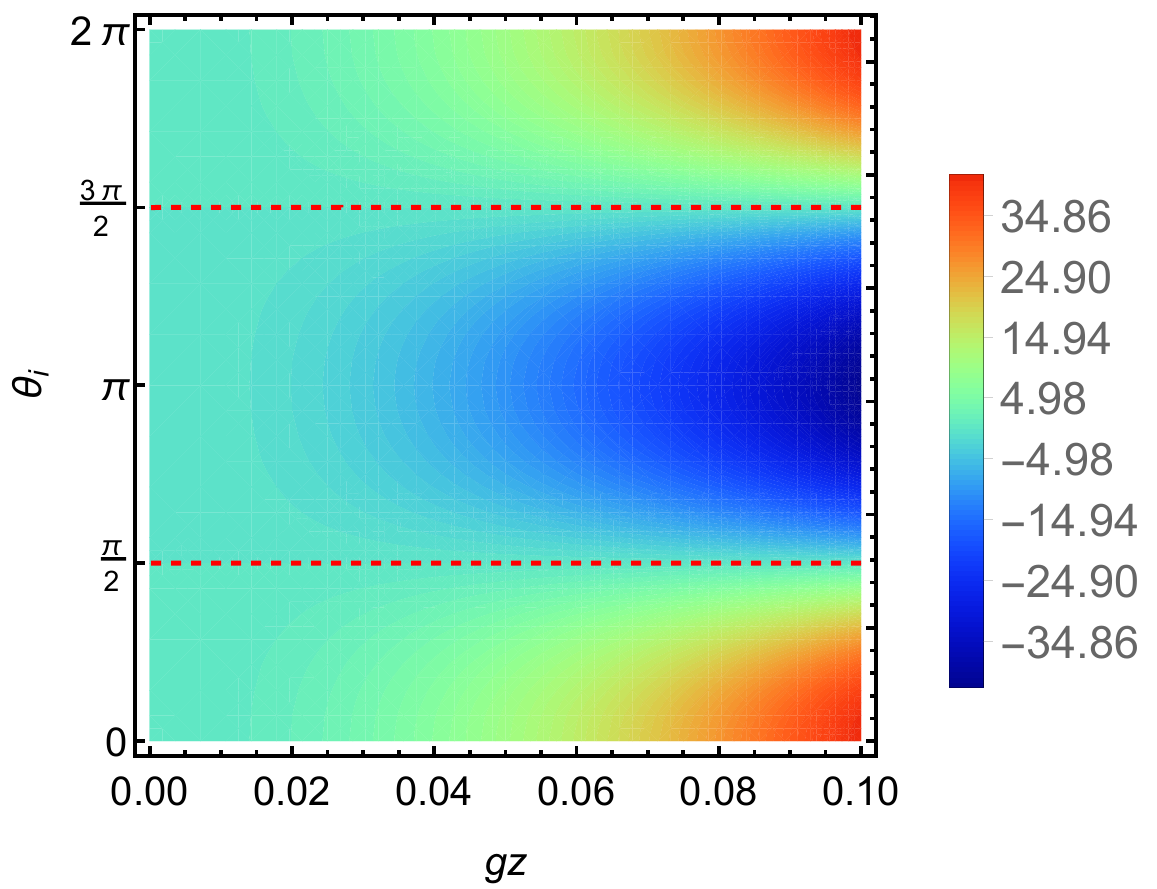} 
\par\end{centering}
} \subfloat[]{\begin{centering}
\includegraphics[scale=0.6]{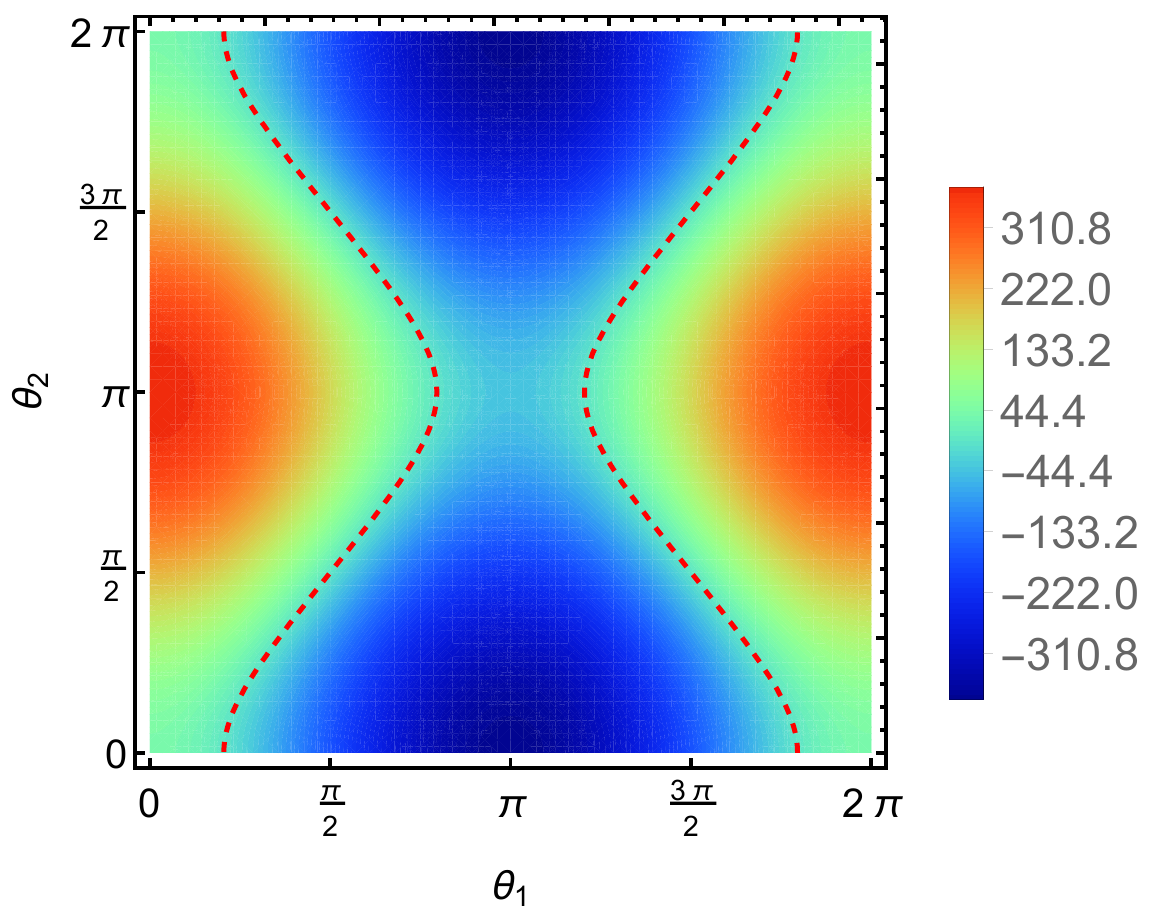}
\par\end{centering}
}
\par\end{centering}
\caption{\label{fig:ph-mis}(Color online) Spatial evolution of Zeno parameter
$Z_{j}$ for (a) Stokes, (b) anti-Stokes, and (c)-(d) phonon modes
as a function of the phase mismatch parameter $\theta_{i}$. In (c),
we have shown $Z_{c}\left(\theta_{1}=\theta_{2}\right)$. We have
chosen here coupling coefficients $\chi=10g,\,\Gamma=100g$ with initial
coherent state amplitudes for all modes $\alpha=11,\alpha_{1}=10,\alpha_{2}=9.5,\beta=8,\gamma=0.01,\delta=1$.
We have also used $\text{\ensuremath{\Delta\omega_{S}=\Delta\omega_{A}=10^{-2}g=10\Delta\omega_{D}}}$.
The dashed (red) contour lines represent a crossover between QZE and
QAZE. All the quantities shown here and the rest of the plots are
dimensionless. }
\end{figure*}

\section{Discussion \label{sec:Discussion}}

There are some interesting scenarios to consider for the present study,
namely spontaneous, stimulated, and partially stimulated cases. Specifically, from the expressions of $\mathcal{C}_{j}$, we can easily observe that 
in the spontaneous case, when $\alpha_{i}\neq0,\alpha\neq0$ and $\beta=\gamma=\delta=0$,
$Z_{j}=0\,\forall j\in\left\{ b,c,d\right\} $. Thus, neither QZE
nor QAZE is observed in the spontaneous case. The stimulated case,
i.e., when all coherent modes are initially prepared with non-zero
intensity, will be discussed in detail later. Prior to that, we may
discus the partially stimulated cases, where among Stokes, anti-Stokes
and phonon modes, initial intensity is nonzero for at most two modes
and the same for other modes(s) is zero; of course the pump modes
have nonzero initial intensity. Consider a particular type of stimulated
case, where $\alpha_{i}\neq0,\alpha\neq0$ and $\beta\neq0,\gamma\neq0,\,\delta=0$.
In this case, we obtain $Z_{d}=0\neq Z_{b}=Z_{c}$; while with $\alpha_{i}\neq0,\alpha\neq0$
and $\beta=0,\gamma\neq0,\,\delta\neq0$ we obtain $Z_{d}=-Z_{c}\neq0=Z_{b}$.
Thus, in the first type of partially stimulated case mentioned above, for a particular choice
of phase mismatching and frequency detuning parameters, if we observe
QZE (QAZE) in Stokes mode, we will observe QZE (QAZE) in phonon mode,
too, but the anti-Stokes mode will not show any of the effects. Similarly,
in the second type of partially stimulated case, for a particular
choice of phase mismatching and frequency detuning parameters, if
we observe QZE (QAZE) in anti-Stokes mode, we will observe QAZE (QZE)
in phonon mode, Stokes mode will not show any of the effects. The
other possibility of partially stimulated process with $\alpha_{i}\neq0,\alpha\neq0$
and $\beta\neq0,\gamma=0,\,\delta\neq0$ leads to a trivial case where
$Z_{j}=0$ as $\mathcal{C}_{j}\propto\left|\gamma\right|\,\forall j\in\left\{ b,c,d\right\} $.
There are a couple of other possibilities where two of the parameters
$\alpha,\beta,\gamma$ are simultaneously zero. It is easy to see
that neither QZE nor QAZE will be observed in those cases. Specifically,
for $\beta=0,\gamma=0,\,\delta\neq0$ and $\beta\neq0,\gamma=0,\,\delta=0$,
we would not obtain QZE or QAZE in any mode as $\gamma=0$ would ensure
that $Z_{j}=0\,\forall j\in\left\{ b,c,d\right\} $. The same situation
will arise in the case $\beta=0,\gamma\neq0,\,\delta=0$ as $\mathcal{C}_{j}$
will vanish $\forall j\in\left\{ b,c,d\right\}$. 

From Eqs. (\ref{eq:Zb})-(\ref{eq:Zd}), we can verify that $\mathcal{C}_{j}\propto2\Gamma\left(\left|\alpha_{1}\right|^{2}+\left|\alpha_{2}\right|^{2}+1\right)\left|\alpha\right|\left|\gamma\right|\,\forall j\in\left\{ b,c,d\right\} $
which is always positive and increases with coupling between the probe
and system as well as initial pump and probe intensities and phonon
numbers. Additionally, we can verify that $\mathcal{C}_{b}\propto g\left|\beta\right|$
and $\mathcal{C}_{d}\propto\chi\left|\delta\right|$. Thus, we can
conclude that the coefficients $\mathcal{C}_{j}$ can only alter the
depth of Zeno parameters by a scaling factor, but cannot induce a transition
from QZE to QAZE or vice-versa. Therefore, we can summarize that the
occurrence of QZE or QAZE would depend only on the detuning parameters
and phase mismatches, i.e., we can safely restrict our discussion
on the possibility of observing QZE and QAZE to a situation where
the parametric dependence of the relevant Zeno parameters is viewed
as $Z_{b}\left(\text{\textgreek{D}}\omega_{D},\text{\textgreek{D}}\omega_{S},\theta_{2}\right)$
and $Z_{d}\left(\text{\textgreek{D}}\omega_{D},\text{\textgreek{D}}\omega_{A},\theta_{1}\right)$. This analytic result helps us to clearly visualize the interplay between QZE and QAZE, and it also answers: What controls the dynamics of QZE and QAZE and which physical parameters may cause transition from one of them to the other? 
Before we discuss it further, it is worth discussing the behavior
for resonant hyper-Raman system waveguide.

\begin{figure}
\begin{centering}
\subfloat[]{
\includegraphics[scale=0.55]{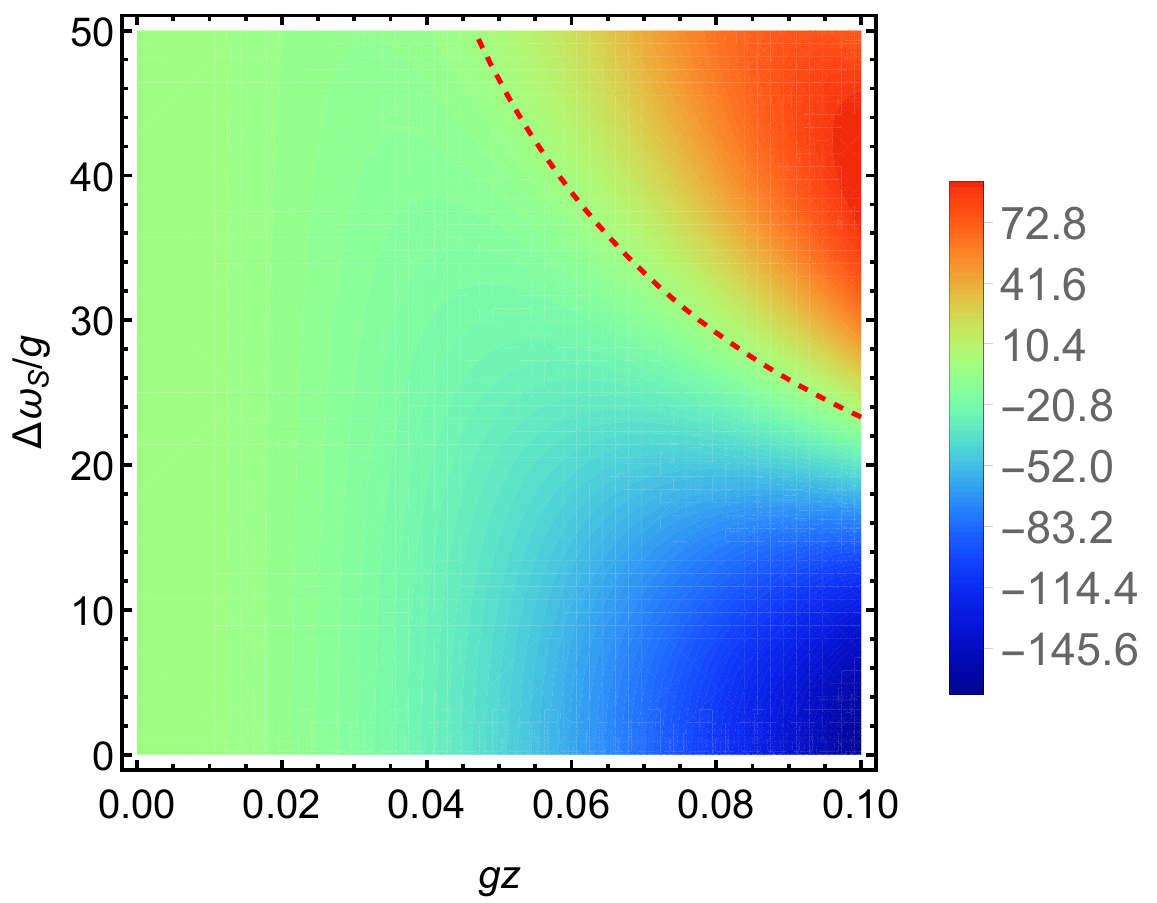}
}\\
 \subfloat[]{
\includegraphics[scale=0.55]{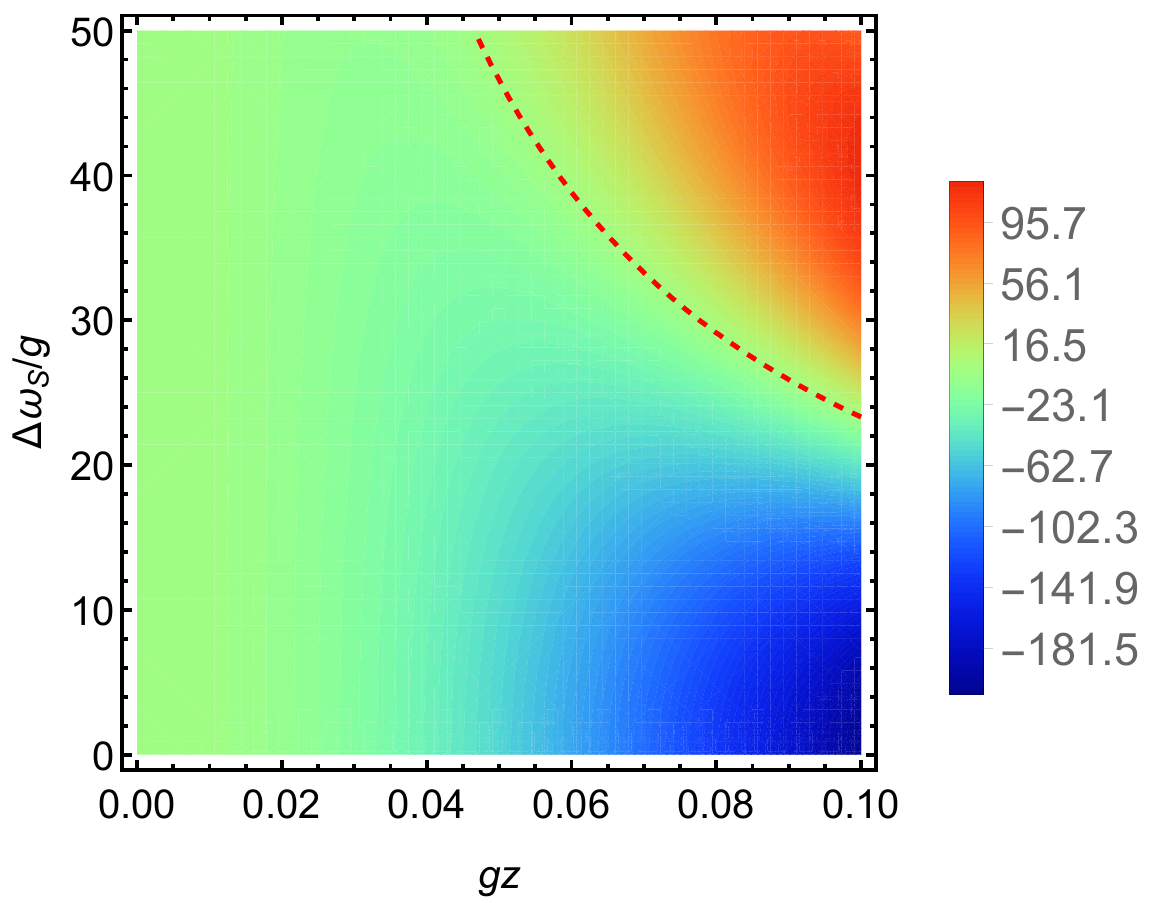}
}\\
 \subfloat[]{
\includegraphics[scale=0.55]{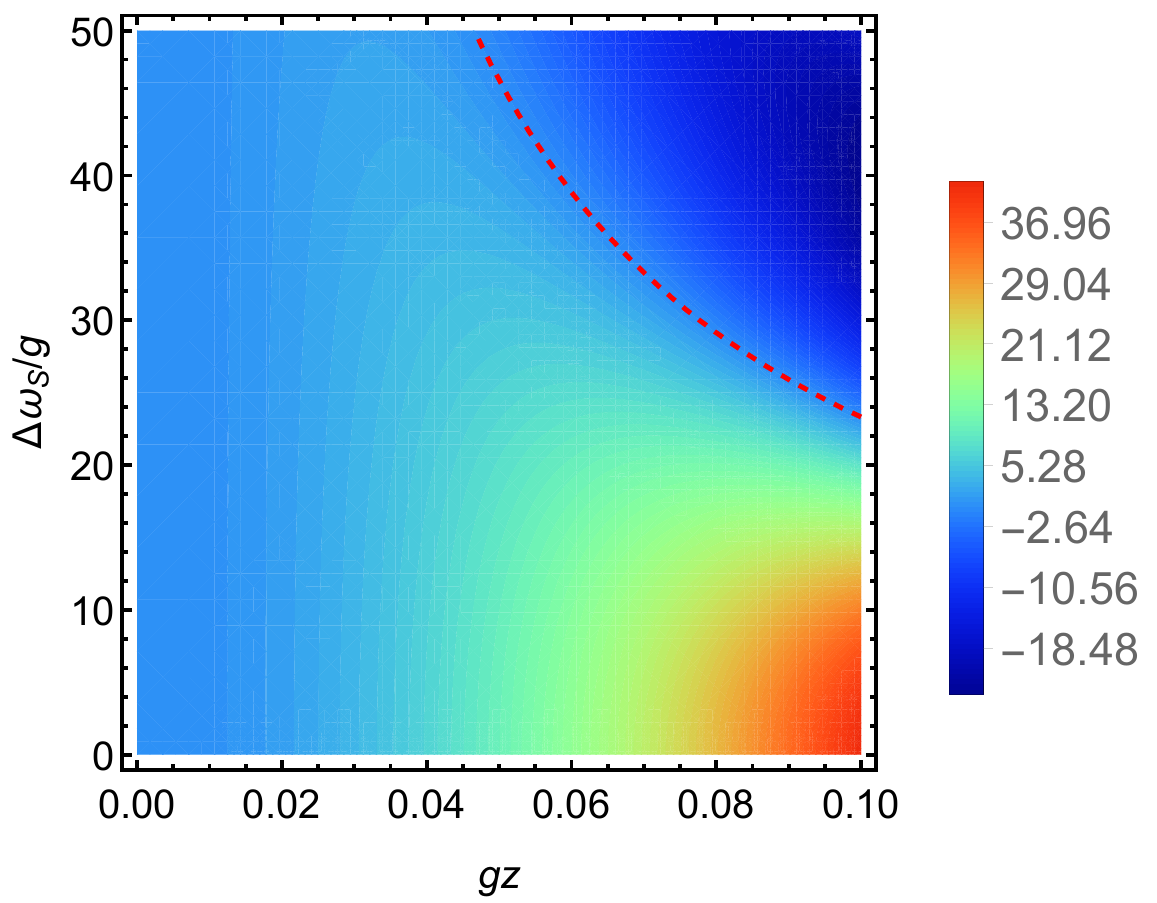}
} 
\par\end{centering}
\caption{\label{fig:det-gz}(Color online) Spatial evolution of Zeno parameter
$Z_{j}$ as a function of frequency detuning parameter for (a) Stokes,
(b) anti-Stokes, and (c) phonon modes considering both phase mismatch
parameters $\theta_{i}=0$ with $\Delta\omega_{D}=10^{-3}g$ and $\Delta\omega_{A}=\Delta\omega_{S}$.
The rest of the parameters are same as in the previous figure.}
\end{figure}

Considering the resonant condition that $\text{\textgreek{D}}\omega_{D}=\text{\textgreek{D}}\omega_{S}=\text{\textgreek{D}}\omega_{A}=0$
we can obtain from Eqs. (\ref{eq:Zb})-(\ref{eq:Zc}), in the limits
of frequency detunings tending to zero, that 
\begin{equation}
\left(Z_{b}\right)_{{\rm R}}=-\frac{1}{2}\mathcal{C}_{b}z^{2}\cos\theta_{2}\label{eq:zb-r}
\end{equation}
and 
\begin{equation}
\left(Z_{d}\right)_{{\rm R}}=-\frac{1}{2}\mathcal{C}_{d}z^{2}\cos\theta_{1},\label{eq:zd-r}
\end{equation}
where ${\rm R}$ corresponds to the resonant condition. Notice that
phase difference parameters are significant in the presence of QZE
and QAZE as $\left(Z_{j}\right)_{{\rm R}}\propto-\cos\theta_{i}$,
while the depth of the Zeno parameters increases with $\mathcal{C}_{j}$
and propagation length $z$. We know that $\cos\theta_{i}>0\,\forall\theta_{i}\in\left\{ \left[0,\frac{\pi}{2}\right]\cup\left[\frac{3\pi}{2},2\pi\right]\right\} $
and $\cos\theta_{i}<0\,\forall\theta_{i}\in\left\{ \left[\frac{\pi}{2},\frac{3\pi}{2}\right]\right\} $.
Thus, QAZE (QZE) is observed in both Stokes and anti-Stokes modes
when $\cos\theta_{i}$ is negative (positive). In this case, $\left(Z_{c}\right)_{{\rm R}}=-\left(\mathcal{C}_{b}\cos\theta_{2}-\mathcal{C}_{d}\cos\theta_{1}\right)z^{2}$,
which has dependence on both phase difference parameters. If we consider $\theta_{1}=\theta_{2}$, this will reduce
to $\left.\left(Z_{c}\right)_{{\rm R}}\right|_{\theta_{1}=\theta_{2}}=-\left(\mathcal{C}_{b}-\mathcal{C}_{d}\right)z^{2}\cos\theta_{i}$,
and thus, in this special case, QZE (QAZE) would be observed in phonon mode
if $\frac{g}{\chi}>\frac{\left|\delta\right|}{\left|\beta\right|}$
$\left(\frac{g}{\chi}<\frac{\left|\delta\right|}{\left|\beta\right|}\right)$.
Interestingly, most of the observations made here regarding phase
mismatch parameters are also applicable in off-resonant hyper-Raman
case with small frequency detunings (cf. Fig. \ref{fig:ph-mis}). 

Consider a special case of off-resonant hyper-Raman process in the
system waveguide, when $\text{\textgreek{D}}\omega_{D}=\text{\textgreek{D}}\omega_{S}=-\text{\textgreek{D}}\omega_{A}$
corresponding to phonon excitation. {Here, $\text{\textgreek{D}}\omega_{S}=-\text{\textgreek{D}}\omega_{A}$
ensures the phonon excitation as $2\omega_{c}=\omega_{d}-\omega_{b}$.
}In this case, Eqs. (\ref{eq:Zb})-(\ref{eq:Zc}) can be simplified
to obtain 
\begin{equation}
\begin{array}{lcl}
Z_{j} & = & -\frac{1}{2}\mathcal{C}_{j}z^{2}\cos\left(\text{\textgreek{D}}\omega_{D}z+(-1)^{i}\theta_{i}\right){\rm sinc}^{2}\left(\frac{\text{\textgreek{D}}\omega_{D}z}{2}\right),\end{array}\label{eq:zj-PhEx}
\end{equation}
where $i=1$ for $j=d$ and $i=2$ for $j=b$. It clearly shows that
the general solution used here is modulated with sinc function of
the frequency detuning parameter \cite{off-res-Raman,off-res-RamanPh}.
Consequently, the present solution and corresponding results are applicable
to a relatively long length of the optical coupler in comparison
to the corresponding short-length solution. Notice in Eq. (\ref{eq:zj-PhEx})
that $Z_{j}\propto\cos\left(\text{\textgreek{D}}\omega_{D}z+(-1)^{j}\theta_{i}\right)$, 
and thus a crossover between QZE and QAZE can be attained by controlling
the frequency detuning $\text{\textgreek{D}}\omega_{D}$. Therefore,
if we assume phase mismatch parameter $\theta_{i}=0$, we can conclude
that $Z_{j}$ is negative (i.e., QZE is observed) only when $\text{\textgreek{D}}\omega_{D}z\in\left\{ \left[0,\frac{\pi}{2}\right]\cup\left[\frac{3\pi}{2},2\pi\right]\right\} $,
otherwise QAZE is observed.

\begin{figure*}
\begin{centering}
\subfloat[]{\begin{centering}
\includegraphics[scale=0.6]{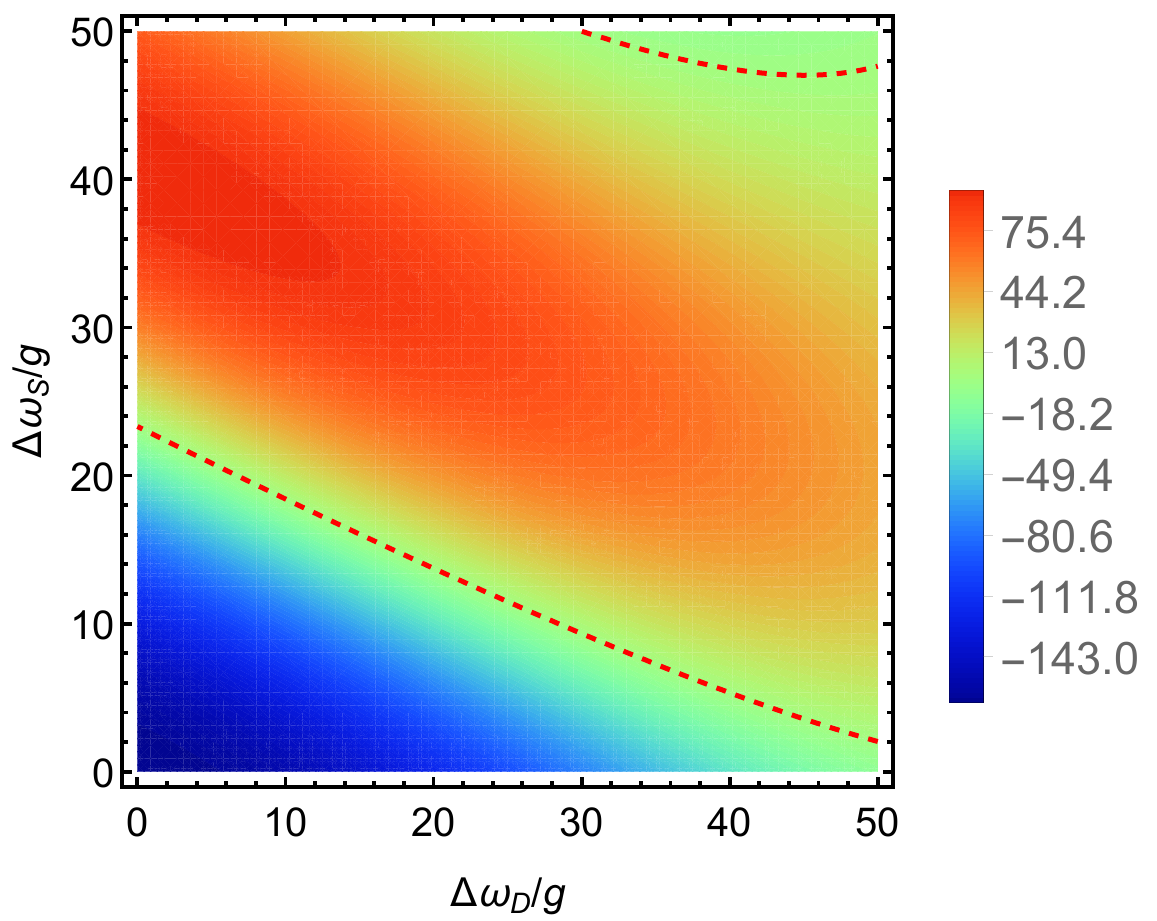}
\par\end{centering}
} \subfloat[]{\begin{centering}
\includegraphics[scale=0.6]{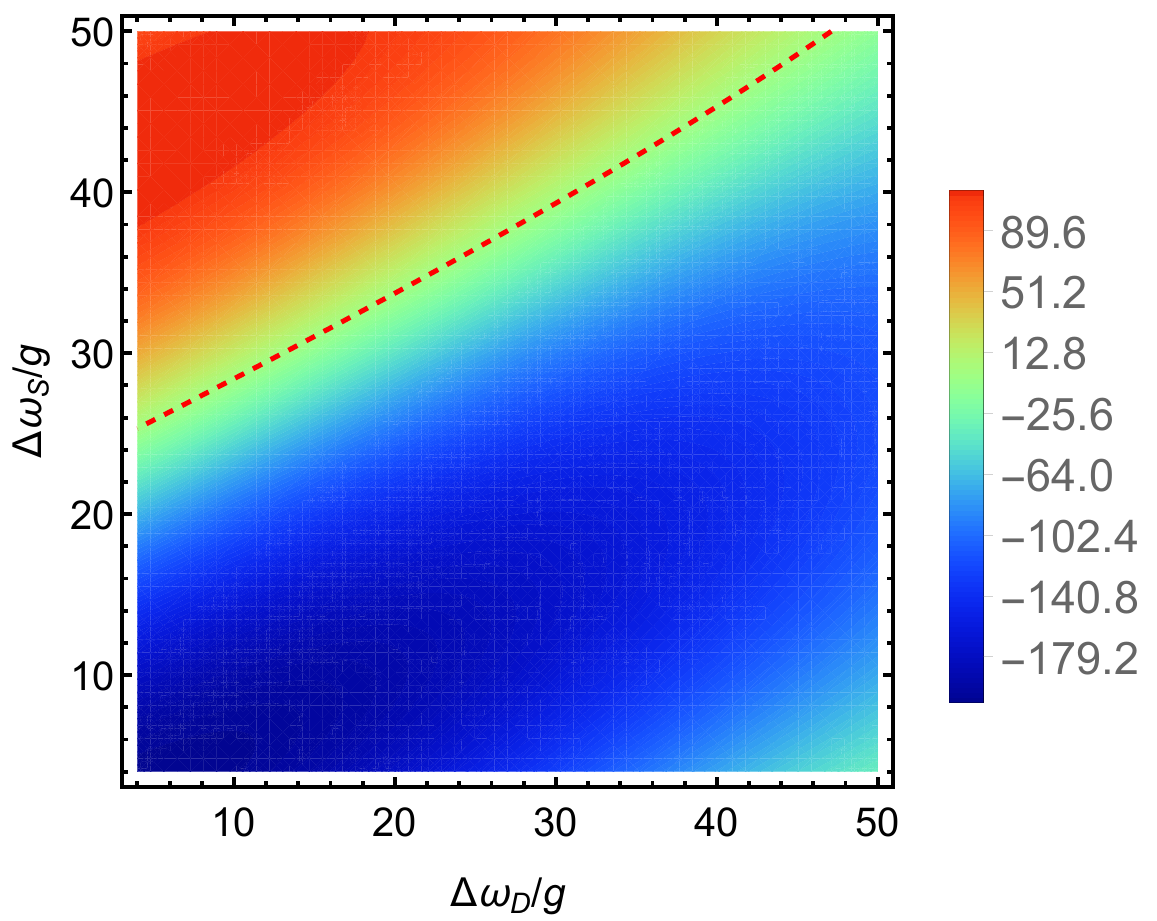}
\par\end{centering}
}
\par\end{centering}
\begin{centering}
\subfloat[]{\begin{centering}
\includegraphics[scale=0.6]{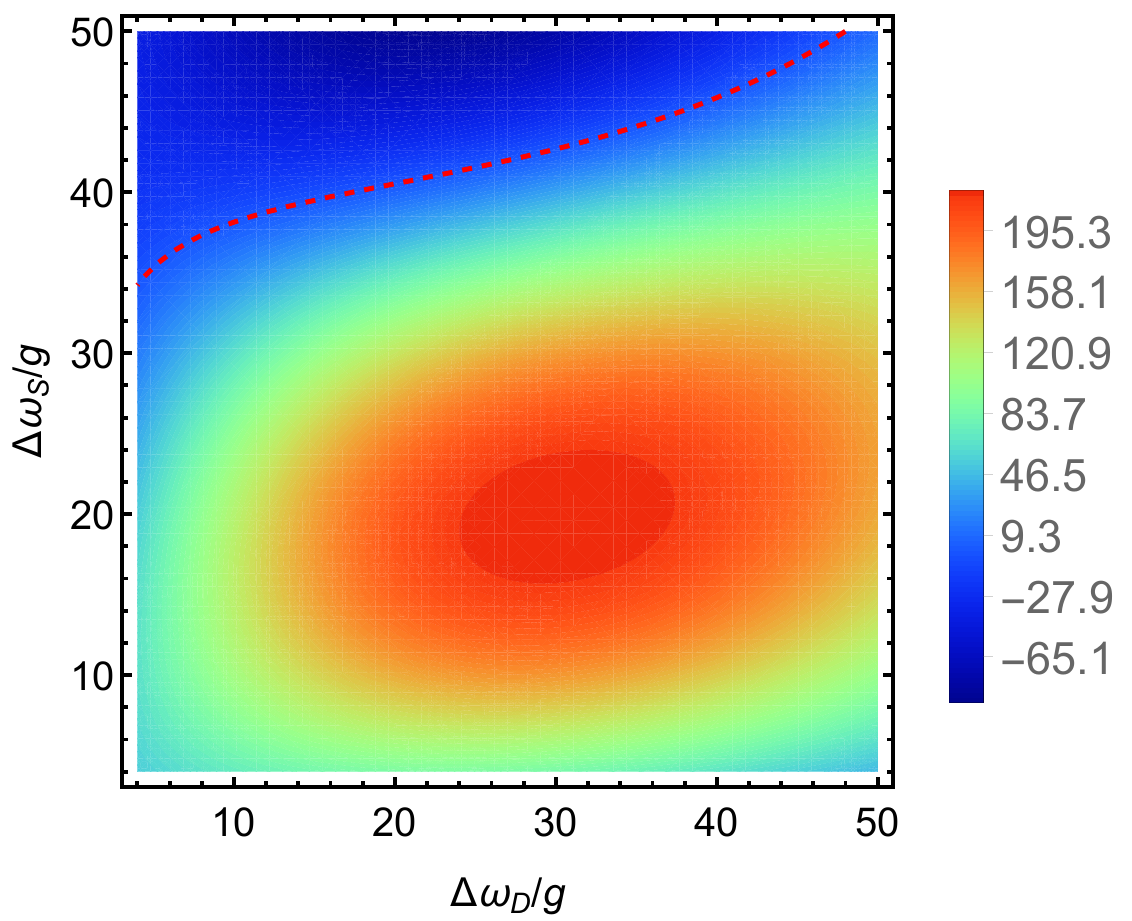}
\par\end{centering}
} \subfloat[]{\begin{centering}
\includegraphics[scale=0.6]{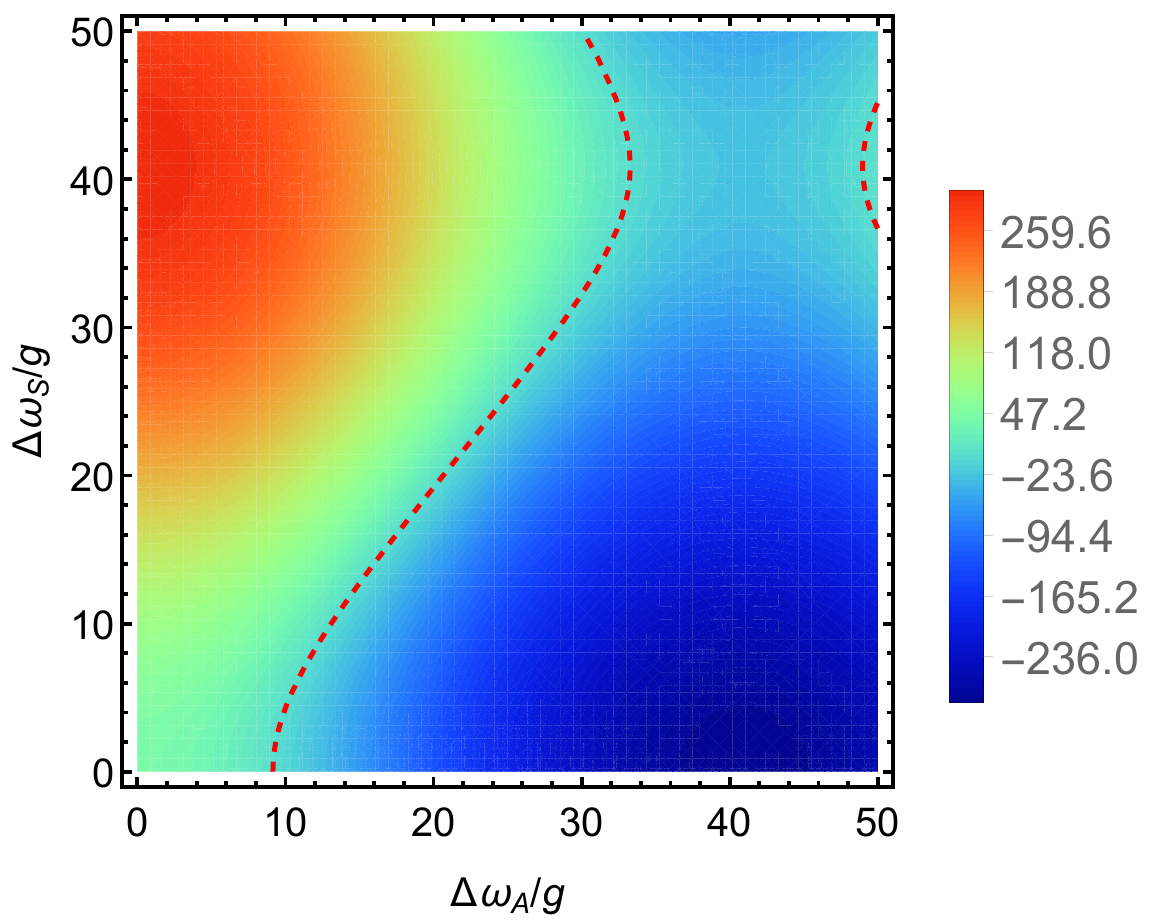}
\par\end{centering}
}
\par\end{centering}
\caption{\label{fig:det}(Color online) Variation of Zeno parameters as a function
of frequency detuning parameters for (a) Stokes $Z_{b}\left(\Delta\omega_{S},\Delta\omega_{D}\right)$,
(b) anti-Stokes $Z_{d}\left(\Delta\omega_{A}=\Delta\omega_{S},\Delta\omega_{D}\right)$,
(c) phonon $Z_{c}\left(\Delta\omega_{A}=\Delta\omega_{S},\Delta\omega_{D}\right)$,
and (c) phonon $Z_{c}\left(\Delta\omega_{S},\Delta\omega_{A},\Delta\omega_{D}=10^{-3}g\right)$
modes considering both phase mismatch parameters $\theta_{i}=0$ and
$gz=0.1$. The rest of the parameters are same as the previous figure.}
\end{figure*}

To further stress on this point, we have shown variation of the Zeno
parameters with frequency detuning parameters in Fig. \ref{fig:det-gz},
where we can clearly see that a transition from QZE (QAZE) to QAZE
(QZE) can be caused in both optical modes (phonon mode) for large
values of frequency detuning after traversing sufficiently through
the optical coupler. {In all the plots, we have considered
$\text{\textgreek{D}}\omega_{S}=\text{\textgreek{D}}\omega_{A}$ (unless
stated otherwise) corresponding to the conservation of radiation energy
as $2\left(\omega_{a_{1}}+\omega_{a_{2}}\right)=\omega_{b}+\omega_{d}$.}

Taking into consideration both frequency detuning parameters together
(considering there is no phase mismatch), a transition from QZE to
QAZE in Stokes mode can be induced for smaller values of the detuning
parameter than that shown in Fig. \ref{fig:det-gz} (a) by increasing
the other detuning parameter (shown in Fig. \ref{fig:det} (a)). In
contrast, QZE for anti-Stokes mode can be maintained for larger values
of frequency detuning in anti-Stokes generation by increasing the
value of frequency detuning between the pump modes of the system and probe
waveguides (cf. Figs. \ref{fig:det-gz} (b) and \ref{fig:det} (b)).\textcolor{blue}{{}
}Similarly, QAZE can be made more dominant by increasing frequency
detuning between the pump modes of the system and probe as shown in Fig.
\ref{fig:det} (c). Interestingly, frequency detuning in Stokes and
anti-Stokes generation processes have starkly opposite effects on
the Zeno parameter as the former increases it while the latter decreases
it (cf. Fig. \ref{fig:det} (d)).

Further extension of the present results with phonon mode initially
coherent to the chaotic phonon mode \cite{off-res-Raman} gives that
all the Zeno parameters become zero. Our results can also be used
to deduce the presence of QZE and QAZE for degenerate hyper-Raman
and Raman system waveguides from Eqs. (\ref{eq:Zb})-(\ref{eq:Zc})
by considering $\alpha_{2}=\alpha_{1}$ and $\alpha_{2}=0$, respectively.
Therefore, we can conclude from the reduced results in those cases
that the presence of QZE and QAZE in the parametric space will remain
unchanged, though some changes in the depth of the Zeno parameter is
expected due to changes in the scaling factors $\mathcal{C}_{j}$.

\section{Conclusion \label{sec:Conclusion}}

The dynamics of a nonlinear waveguide operating under
hyper-Raman process is obtained in terms of the spatial evolution of the photon
and phonon numbers of the Stokes, anti-Stokes, and vibration modes.
We subsequently consider that this waveguide (referred to as a system)
is constantly gazed upon by a probe waveguide interacting with the
non-degenerate pump modes of the hyper-Raman process. We obtain the
dynamics of the system waveguide in this combined system-probe optical
coupler in a completely quantum treatment by solving the Heisenberg's
equations of motion for the corresponding momentum operator. These
two cases allow us to quantify the effect of the presence of the probe
waveguide on the system waveguide as Zeno parameter. Specifically,
if the presence of the probe is found to enhance (suppress) the generation
of bosons in Stokes, anti-Stokes, and phonon modes it is referred to as QAZE
(QZE). 

The conservation of Stokes--anti-Stokes photon and phonon numbers
is reflected in the relation between the Zeno parameters for the concerned
modes. To be specific, the Zeno parameter for the phonon mode is the
difference between that of the Stokes and anti-Stokes modes. The present
study allows us to conclude that both QZE and QAZE disappear in spontaneous
and some of the partially stimulated cases. Interestingly, the Zeno
parameter depends on the intensities of pump, probe, and phonon modes
as well as the system-probe coupling strength. However, none of these
parameters can cause a crossover from QZE to QAZE and vice versa, as
they always remain positive and thus can only alter the depth of the
Zeno parameter. A similar effect was shown by the spatial evolution
for the small values of frequency detuning parameters.

Interestingly, the hyper-Raman based optical coupler both at resonance
and off-resonance shown dependence on phase mismatch parameters in
Stokes and anti-Stokes generation processes. Similarly, the frequency
detuning parameters offer another set of control parameters to induce
a transition between QZE and QAZE.  In short, we have analytically obtained solution of an interesting question: Which physical parameters controls the dynamics of QZE and QAZE and which of them can cause a transition from QZE to QAZE and vice versa.

Due to a general nature of the system under consideration, namely
non-degenerate hyper-Raman process, the obtained results can be reduced
to corresponding Raman and degenerate hyper-Raman processes. Specifically,
the dependence on all the parameters and the nature of the Zeno parameter
in these special cases is expected to be similar to the present case with
only a scaling factor. We conclude the present work in hope that this work
focused on foundationally relevant topic will lead to applications
in counterfactual quantum communication and computation as the physical system considered here and all its special cases are physically realizable using current technologies available in the domain of integrated optics as well as conventional bulk optics.

\textbf{Acknowledgments: }AP acknowledges the support from Interdisciplinary
Cyber Physical Systems (ICPS) programme of the Department of Science
and Technology (DST), India, Grant No.: DST/ICPS/QuST/Theme-1/2019/14.
KT acknowledges the financial support from the Operational Programme
Research, Development and Education - European Regional Development
Fund project no. CZ.02.1.01/0.0/0.0/16\_019/0000754 of the Ministry
of Education, Youth and Sports of the Czech Republic.

\appendix

\section{Mathematical details of the solution \label{sec:app-A}}

The Heisenberg's equations of motion for all six modes in the probe
and system waveguides can be obtained as 
\begin{equation}
\begin{array}{lcl}
\overset{.}{a_{p}}\left(z\right) & = & i\left(\omega_{p}a_{p}+\Gamma a_{1}a_{2}\right),\\
\overset{.}{a_{1}}\left(z\right) & = & i\left(\omega_{a_{1}}a_{1}+ga_{2}^{\dagger}bc+\chi a_{2}^{\dagger}c^{\dagger}d+\Gamma a_{p}a_{2}^{\dagger}\right),\\
\overset{.}{a_{2}}\left(z\right) & = & i\left(\omega_{a_{2}}a_{2}+ga_{1}^{\dagger}bc+\chi a_{1}^{\dagger}c^{\dagger}d+\Gamma a_{p}a_{1}^{\dagger}\right),\\
\overset{.}{b}\left(z\right) & = & i\left(\omega_{b}b+ga_{1}a_{2}c^{\dagger}\right),\\
\overset{.}{c}\left(z\right) & = & i\left(\omega_{c}c+ga_{1}a_{2}b^{\dagger}+\chi a_{1}^{\dagger}a_{2}^{\dagger}d\right),\\
\overset{.}{d}\left(z\right) & = & i\left(\omega_{d}d+\chi a_{1}a_{2}c\right).
\end{array}\label{eq:HesE}
\end{equation}
This set of coupled operator differential equations is not exactly
solvable in the closed analytic form. A perturbative solution of the
coupled differential equations (\ref{eq:HesE}) is obtained up to
the quardatic terms in the interaction constants ($g,$$\chi$, and
$\Gamma$). Using the obtained spatial evolution of all the field
and phonon modes in the closed analytic form, we obtained the number
operators for Stokes, phonon, and anti-Stokes modes as 
\begin{equation}
\begin{array}{lcl}
\left\langle N_{b}\right\rangle  & = & \left|\beta\right|^{2}+\left|j_{2}\right|^{2}\left|\alpha_{1}\right|^{2}\left|\alpha_{2}\right|^{2}\left(\left|\gamma\right|^{2}+1\right)+\left[\left\{ j_{1}j_{2}^{\star}\alpha_{1}^{\star}\alpha_{2}^{\star}\right.\right.\\
&\times&\beta\gamma+j_{1}j_{3}^{\star}\alpha_{1}^{\star2}\alpha_{2}^{\star2}\beta\delta+j_{1}j_{4}^{\star}\left(\left|\alpha_{1}\right|^{2}+1\right)\beta\gamma^{2}\delta^{\star}\\
 & + & j_{1}j_{5}^{\star}\left|\alpha_{2}\right|^{2}\beta\gamma^{2}\delta^{\star}+j_{1}j_{6}^{\star}\left(\left|\alpha_{1}\right|^{2}+1\right)\alpha^{\star}\beta\gamma\\
 &+&j_{1}j_{7}^{\star}\left|\alpha_{2}\right|^{2}\alpha^{\star}\beta\gamma+j_{1}j_{8}^{\star}\left|\alpha_{1}\right|^{2}\left|\alpha_{2}\right|^{2}\left|\beta\right|^{2}+j_{1}j_{9}^{\star}\left|\beta\right|^{2}\\
 & \times& \left.\left.\left(\left|\alpha_{1}\right|^{2}+1\right)\left|\gamma\right|^{2}+j_{1}j_{10}^{\star}\left|\alpha_{2}\right|^{2}\left|\beta\right|^{2}\left|\gamma\right|^{2}+{\rm c.c.}\right\} \right],
\end{array}\label{eq:nb}
\end{equation}
\begin{equation}
\begin{array}{lcl}
\left\langle N_{c}\right\rangle  & = & \left|\gamma\right|^{2}+\left|k_{2}\right|^{2}\left|\alpha_{1}\right|^{2}\left|\alpha_{2}\right|^{2}\left(\left|\beta\right|^{2}+1\right)+\left|k_{3}\right|^{2}\\
&\times&\left(\left|\alpha_{1}\right|^{2}+1\right)\left(\left|\alpha_{2}\right|^{2}+1\right)\left|\delta\right|^{2}+\left[\left\{ k_{1}k_{2}^{\star}\alpha_{1}^{\star}\alpha_{2}^{\star}\beta\gamma\right.\right.\\
 & + & k_{1}k_{3}^{\star}\alpha_{1}\alpha_{2}\gamma\delta^{\star}+k_{1}k_{4}^{\star}\left(\left|\alpha_{1}\right|^{2}+1\right)\alpha^{\star}\beta\gamma\\
 & + &k_{1}k_{5}^{\star}\left|\alpha_{2}\right|^{2}\alpha^{\star}\beta\gamma+k_{1}k_{6}^{\star}\left|\alpha_{1}\right|^{2}\alpha\gamma\delta^{\star}\\
 & + & k_{1}k_{7}^{\star}\left(\left|\alpha_{2}\right|^{2}+1\right)\alpha\gamma\delta^{\star}+k_{2}k_{3}^{\star}\alpha_{1}^{2}\alpha_{2}^{2}\beta^{\star}\delta^{\star}\\
 & + &k_{1}k_{8}^{\star}\left|\alpha_{1}\right|^{2}\left|\alpha_{2}\right|^{2}\left|\gamma\right|^{2}+k_{1}k_{9}^{\star}\left(\left|\alpha_{1}\right|^{2}+1\right)\\
 &\times& \left|\beta\right|^{2}\left|\gamma\right|^{2}
  +  k_{1}k_{10}^{\star}\left|\alpha_{2}\right|^{2}\left|\beta\right|^{2}\left|\gamma\right|^{2}+k_{1}k_{11}^{\star}\left|\alpha_{1}\right|^{2}\\
 &\times& \left|\gamma\right|^{2}\left|\delta\right|^{2}
  + k_{1}k_{12}^{\star}\left(\left|\alpha_{2}\right|^{2}+1\right)\left|\gamma\right|^{2}\left|\delta\right|^{2}\\
 & + & \left.\left.k_{1}k_{13}^{\star}\left|\alpha_{1}\right|^{2}\left|\alpha_{2}\right|^{2}\left|\gamma\right|^{2}\right\} +{\rm c.c.}\right],
\end{array}\label{eq:nc}
\end{equation}
and
\begin{equation}
\begin{array}{lcl}
\left\langle N_{d}\right\rangle  & = & \left|\delta\right|^{2}+\left|l_{2}\right|^{2}\left|\alpha_{1}\right|^{2}\left|\alpha_{2}\right|^{2}\left|\gamma\right|^{2}+\left[\left\{ l_{1}l_{2}^{\star}\alpha_{1}^{\star}\alpha_{2}^{\star}\gamma^{\star}\delta\right.\right.\\
&+& l_{1}l_{3}^{\star}\left(\left|\alpha_{1}\right|^{2}+1\right)\beta^{\star}\gamma^{\star2}\delta+l_{1}l_{4}^{\star}\left|\alpha_{2}\right|^{2}\beta^{\star}\gamma^{\star2}\delta\\
 & + & l_{1}l_{5}^{\star}\alpha_{1}^{\star2}\alpha_{2}^{\star2}\beta\delta+l_{1}l_{6}^{\star}\left(\left|\alpha_{1}\right|^{2}+1\right)\alpha^{\star}\gamma^{\star}\delta\\
 &+&l_{1}l_{7}^{\star}\left|\alpha_{2}\right|^{2}\alpha^{\star}\gamma^{\star}\delta+l_{1}l_{8}^{\star}\left(\left|\alpha_{1}\right|^{2}+1\right)\left|\gamma\right|^{2}\left|\delta\right|^{2}\\
 & + & l_{1}l_{9}^{\star}\left|\alpha_{2}\right|^{2}\left|\gamma\right|^{2}\left|\delta\right|^{2}+l_{1}l_{10}^{*}\left(\left|\alpha_{1}\right|^{2}+1\right)\\
 &\times& \left.\left.\left(\left|\alpha_{2}\right|^{2}+1\right)\left|\delta\right|^{2}\right\} +{\rm c.c.}\right],
\end{array}\label{eq:nd}
\end{equation}
respectively. The functional form of coefficients of complex amplitude
parameters are

\begin{equation}
\begin{array}{lcl}
j_{1} & = & e^{iz\omega_{b}},\\
\frac{j_{2}}{j_{1}} & = & \frac{g\left(1-e^{-iz\Delta\omega_{S}}\right)}{\Delta\omega_{S}},\\
\frac{j_{3}}{j_{1}} & = & \frac{g\chi\left(\Delta\omega_{A}-\Delta\omega_{1}e^{-iz\Delta\omega_{S}}-\Delta\omega_{S}e^{iz\Delta\omega_{1}}\right)}{\Delta\omega_{A}\Delta\omega_{1}\Delta\omega_{S}},\\
\frac{j_{4}}{j_{1}} & = & \frac{j_{5}}{j_{1}}=\frac{g\chi\left(\Delta\omega_{A}+\Delta\omega_{S}e^{-iz\Delta\omega_{2}}-\Delta\omega_{2}e^{-iz\Delta\omega_{S}}\right)}{\Delta\omega_{A}\Delta\omega_{S}\Delta\omega_{2}},\\
\frac{j_{6}}{j_{1}} & = & \frac{j_{7}}{j_{1}}=\frac{g\Gamma\left(\Delta\omega_{D}+\Delta\omega_{S}e^{-iz\Delta\omega_{3}}-\Delta\omega_{3}e^{-iz\Delta\omega_{S}}\right)}{\Delta\omega_{D}\Delta\omega_{S}\Delta\omega_{3}},\\
\frac{j_{8}}{j_{1}} & = & -\frac{j_{9}}{j_{1}}=-\frac{j_{10}}{j_{1}}=\frac{g^{2}\left(1-e^{-iz\Delta\omega_{S}}-i\Delta\omega_{S}z\right)}{\Delta\omega_{S}^{2}},
\end{array}
\end{equation}
 
\begin{equation}
{\normalcolor {\color{blue}{\normalcolor {\color{blue}{\normalcolor \begin{array}{lcl}
k_{1} & = & e^{iz\omega_{c}},\\
\frac{k_{2}}{k_{1}} & = & \frac{g\left(1-e^{-iz\Delta\omega_{S}}\right)}{\Delta\omega_{S}},\\
\frac{k_{3}}{k_{1}} & = & \frac{\chi\left(1-e^{-iz\Delta\omega_{A}}\right)}{\Delta\omega_{A}},\\
\frac{k_{4}}{k_{1}} & = & \frac{k_{5}}{k_{1}}=\frac{g\Gamma\left(\Delta\omega_{D}+\Delta\omega_{S}e^{-iz\Delta\omega_{3}}-\Delta\omega_{3}e^{-iz\Delta\omega_{S}}\right)}{\Delta\omega_{D}\Delta\omega_{S}\Delta\omega_{3}},\\
\frac{k_{6}}{k_{1}} & = & \frac{k_{7}}{k_{1}}=\frac{\Gamma\chi\left(-\Delta\omega_{D}+\Delta\omega_{A}e^{-iz\Delta\omega_{4}}-\Delta\omega_{4}e^{-iz\Delta\omega_{A}}\right)}{\Delta\omega_{A}\Delta\omega_{D}\Delta\omega_{4}},\\
\frac{k_{8}}{k_{1}} & = & -\frac{k_{11}}{k_{1}}=-\frac{k_{12}}{k_{1}}=-\frac{\chi^{2}\left(1-e^{-iz\Delta\omega_{A}}-i\Delta\omega_{A}z\right)}{\Delta\omega_{A}^{2}},\\
\frac{k_{9}}{k_{1}} & = & \frac{k_{10}}{k_{1}}=-\frac{k_{13}}{k_{1}}=-\frac{g^{2}\left(1-e^{iz\Delta\omega_{S}}-i\Delta\omega_{S}z\right)}{\Delta\omega_{S}^{2}},
\end{array}}}}}}
\end{equation}
and
\begin{equation}
\begin{array}{lcl}
l_{1} & = & e^{iz\omega_{d}},\\
\frac{l_{2}}{l_{1}} & = &- \frac{\chi\left(1-e^{iz\Delta\omega_{A}}\right)}{\Delta\omega_{A}},\\
\frac{l_{3}}{l_{1}} & = & \frac{l_{4}}{l_{1}}=\frac{g\chi\left(\Delta\omega_{S}+\Delta\omega_{A}e^{iz\Delta\omega_{2}}-\Delta\omega_{2}e^{iz\Delta\omega_{A}}\right)}{\Delta\omega_{S}\Delta\omega_{A}\Delta\omega_{2}},\\
\frac{l_{5}}{l_{1}} & = & \frac{g\chi\left(\Delta\omega_{S}+\Delta\omega_{1}e^{iz\Delta\omega_{A}}-\Delta\omega_{A}e^{iz\Delta\omega_{1}}\right)}{\Delta\omega_{S}\Delta\omega_{1}\Delta\omega_{A}},\\
\frac{l_{6}}{l_{1}} & = & \frac{l_{7}}{l_{1}}=\frac{\Gamma\chi\left(\Delta\omega_{D}-\Delta\omega_{A}e^{iz\Delta\omega_{4}}+\Delta\omega_{4}e^{iz\Delta\omega_{A}}\right)}{\Delta\omega_{D}\Delta\omega_{A}\Delta\omega_{4}},\\
\frac{l_{8}}{l_{1}} & = & \frac{l_{9}}{l_{1}}=\frac{l_{10}}{l_{1}}=-\frac{\chi^{2}\left(1-e^{-iz\Delta\omega_{A}}+i\Delta\omega_{A}z\right)}{\Delta\omega_{A}^{2}},
\end{array}
\end{equation}
where $\Delta\omega_{1}=\left(2\omega_{a_{1}}+2\omega_{a_{2}}-\omega_{b}-\omega_{d}\right)$,
$\Delta\omega_{2}=\left(\omega_{b}+2\omega_{c}-\omega_{d}\right)$,
$\Delta\omega_{3}=\left(\omega_{b}+\omega_{c}-\omega_{p}\right),$
$\Delta\omega_{4}=\left(\omega_{c}-\omega_{d}+\omega_{p}\right),$ $\Delta\omega_{S}=\left(-\omega_{a_{1}}-\omega_{a_{2}}+\omega_{b}+\omega_{c}\right),$ $\Delta\omega_{A}=\left(\omega_{a_{1}}+\omega_{a_{2}}+\omega_{c}-\omega_{d}\right)$,
and $\Delta\omega_{D}=\left(\omega_{a_{1}}+\omega_{a_{2}}-\omega_{p}\right)$.

\begin{thebibliography}{10}
\bibitem{misra1977zeno}B. Misra and E. C. G. Sudarshan, J. Math.
Phys. \textbf{18}, 756-763 (1977).

\bibitem{Khalfin}L. A. Khalfin, Dokl. Akad. Nauk SSSR \textbf{115},
277 (1957) {[}Sov. Phys. Dokl.\textbf{ 2}, 232 (1958){]}; Zh. Eksp.
Teor. Fiz.\textbf{ 33}, 1371 (1958) {[}Sov. Phys. JETP \textbf{6},1053
(1958){]}; Dokl. Akad. Nauk SSSR \textbf{141}, 599 (1961) {[}Sov.
Phys. Dokl. \textbf{6}, 1010 (1962){]}

\bibitem{venugopalan2007zeno_review}A. Venugopalan, Resonance \textbf{12},
52 (2007).

\bibitem{facchi2001zeno-review}P. Facchi and S. Pascazio, in \textit{Quantum
Zeno and Inverse Quantum Zeno Effects}, edited by E. Wolf, Progress
in Optics Vol. 42 (Elsevier, Amsterdam, 2001), pp. 147-218.

\bibitem{saverio-zeno-in-70-minutes}S. Pascazio, Open Systems and
Information Dynamics \textbf{21}, 1440007 (2014).

\bibitem{chi2-chi1-spie}K. Thapliyal and A. Pathak, Proceedings of
International Conference on Optics and Photonics, February 20-22,
Kolkata, India , Proc. of SPIE \textbf{9654}, 96541F1 (2015).

\bibitem{PT}J. Naikoo, K.Thapliyal, S. Banerjee, and A. Pathak, Phys. 
Rev. A \textbf{99}, 023820 (2019).

\bibitem{NZeno}K. Thapliyal, A. Pathak, and J. Pe$\check{{\rm r}}$ina, Phys. Rev.
A \textbf{93}, 022107 (2016).

\bibitem{saverio-3-manifestation}P. Facchi, and S. Pascazio, Three
different manifestations of the quantum Zeno effect. Lecture notes
in Physics \textbf{622}, 141 (2003).

\bibitem{thun2002zeno-raman}K. Thun, J. Pe$\check{{\rm r}}$ina,
and J. K$\breve{{\rm r}}$epelka, Phys. Lett. A \textbf{299}, 19 (2002).

\bibitem{Rechacek-2001-zrno-coupler}J. ${\rm \check{R}eh\acute{a}\check{c}ek}$,
J. Pe$\check{{\rm r}}$ina, P. Facchi, S. Pascazio, and L. Mi${\rm \check{s}}$ta
Jr, Opt. Spectrosc. \textbf{91}, 501 (2001).

\bibitem{chi2-chi2}L. Mi${\rm \check{s}}$ta Jr, V. Jel${\rm \acute{\imath}}$nek,
J. Reh${\rm \acute{a}}$cek, and J. Pe$\check{{\rm r}}$ina, J. Opt.
B: Quantum Semiclassical Opt. \textbf{2}, 726 (2000).

\bibitem{rehacek2000zeno-coupler}J. ${\rm \check{R}eh\acute{a}\check{c}ek}$,
J. Pe$\check{{\rm r}}$ina, P. Facchi, S. Pascazio, and L. Mi${\rm \check{s}}$ta,
Phys. Rev. A \textbf{62}, 013804 (2000).

\bibitem{down-conversion-perina2}A. Luis and J. Pe$\check{{\rm r}}$ina,
Phys. Rev. Lett. \textbf{76}, 4340 (1996).

\bibitem{parametric-down-conversion-anti-zeno}A. Luis and L. L. S${\rm \acute{a}}$nchez-Soto,
Phys. Rev. A \textbf{57}, 781 (1998).

\bibitem{down-conversion-perina}J. ${\rm \check{R}eh\acute{a}\check{c}ek}$,
J. Pe$\check{{\rm r}}$ina, P. Facchi, S. Pascazio, and L. Mi${\rm \check{s}}$ta,
Phys. Rev. A \textbf{62}, 013804 (2000).

\bibitem{perina-zeno-parametric-dc}J. Pe$\check{{\rm r}}$ina, Phys.
Lett. A \textbf{325}, 16 (2004).

\bibitem{All-optical-zeno-agarwal}G. S. Agarwal and S. P. Tewari,
Phys. Lett. A \textbf{185}, 139 (1994).

\bibitem{kwait-int.-free.measurement-expt.}P. G. Kwiat, A. G. White,
J. R. Mitchell, O. Nairz, G. Weihs, H. Weinfurter, and A. Zeilinger,
Phys. Rev. Lett. \textbf{83}, 4725 (1999).

\bibitem{experimental-zeno-1}W. M. Itano, D. J. Heinzen, J. J. Bollinger,
and D. J. Wineland, Phys. Rev. A \textbf{41}, 2295 (1990).

\bibitem{experimental-zeno-2}M. C. Fischer, B. Guti${\rm \acute{e}}$rrez-Medina,
and M. G. Raizen, Phys. Rev. Lett. \textbf{87}, 040402 (2001).

\bibitem{counterfactual-quantum-computation}O. Hosten, M. T. Rakher,
J. T. Barreiro, N. A. Peters, and P. G. Kwiat, Nature \textbf{439},
949 (2006).

\bibitem{Zubairy1}H. Salih, Z. H. Li, M. Al-Amri, and M. S. Zubairy,
Phys. Rev. Lett. \textbf{110}, 170502 (2013).

\bibitem{zeno-tomography-hradil}P. Facchi, Z. Hradil, G. Krenn, S.
Pascazio, and J. ${\rm \check{R}eh\acute{a}\check{c}ek}$, Phys. Rev.
A \textbf{66}, 012110 (2002).

\bibitem{zeno-tomography2}S. Pascazio, P. Facchi, Z. Hradil, G. Krenn,
and J. Rehacek, Fortschritte der Physik \textbf{49}, 1071 (2001).

\bibitem{Q-com-comp}A. Tavakoli, H. Anwer, A. Hameedi, and M. Bourennane,
Phys. Rev. A \textbf{92}, 012303 (2015).

\bibitem{Decoherence-free}A. Beige, D. Braun, B. Tregenna, and P.
L. Knight, Phys. Rev. Lett. \textbf{85}, 1762 (2000).

\bibitem{coun-comEx}Y. Cao, Y.-H. Li, Z. Cao, J. Yin, Y.-A.
Chen, H.-L. Yin, T.-Y. Chen, X. Ma, C.-Z. Peng, and
J.-W. Pan, Proc. of the Nat. Acad. Sci. \textbf{114}, 4920 (2017).

\bibitem{Zeno-blackhole}H. Nikoli${\rm \acute{c}}$, Phys. Lett.
B \textbf{733}, 6 (2014).

\bibitem{macroscopic-zenoPRL}D. A. Zezyulin, V. V. Konotop, G. Barontini,
and H. Ott, Phys. Rev. Lett. \textbf{109}, 020405 (2012).

\bibitem{obrien1}J. L. O'brien, A. Furusawa, and J. Vu\v{c}kovi\'{c},
Nature Photonics \textbf{3}, 687 (2009).

\bibitem{obrien2}A. Politi, M. J. Cryan, J. G. Rarity,
S. Yu, and J. L. O'brien, Science \textbf{320}, 646
(2008).

\bibitem{kishore2014co-coupler}K. Thapliyal, A. Pathak, B. Sen, and
J. Pe$\check{{\rm r}}$ina, Phys. Rev. A \textbf{90}, 013808 (2014).

\bibitem{kishore2014contra}K. Thapliyal, A. Pathak, B. Sen, and J.
Pe$\check{{\rm r}}$ina, Phys. Lett. A \textbf{378}, 3431-3440 (2014).

\bibitem{bsen1} B Sen and S Mandal, J. Mod. Opt. \textbf{52}, 1789
(2005).

\bibitem{mandal2004co-coupler}S. Mandal and J. Pe$\check{{\rm r}}$ina,
Phys. Lett. A \textbf{328}, 144 (2004).

\bibitem{perina-book}J. Pe$\check{{\rm r}}$ina, Quantum Statistics of Linear and
Nonlinear Optical Phenomena, Kluwer Academic, Dordrecht-Boston (1991).

\bibitem{off-res-Raman} K. Thapliyal and J. Pe$\check{{\rm r}}$ina, Phys. Lett. 
A \textbf{383}, 2011 (2019).

\bibitem{off-res-RamanPh} K. Thapliyal and J. Pe$\check{{\rm r}}$ina, Phys. Scr.
\textbf{95}, 034001 (2020).
\end{thebibliography}
\end{document}